\documentclass[oribibl]{llncs}

\usepackage{algorithm}
\usepackage[noend]{algpseudocode}
\usepackage{amsmath}
\usepackage{amssymb}
\usepackage{array}
\usepackage{caption}
\usepackage{cmap}  
\usepackage{comment}
\usepackage{csquotes}
\usepackage[shortlabels]{enumitem}       
\usepackage{filecontents}
\usepackage{fontenc}  
\usepackage{mathtools}
\usepackage{microtype}
\usepackage[numbers]{natbib}
\usepackage{subcaption}

\DisableLigatures[f]{encoding = *, family = *}

\usepackage[hyperfootnotes=false,pdfpagemode=UseNone,pdfstartview=]{hyperref}

\DeclarePairedDelimiter{\p}{\lparen}{\rparen\nonscript\!\mathinner{}}
\DeclarePairedDelimiter{\abs}{\lvert}{\rvert}

\DeclarePairedDelimiter{\bk}{\lbrack}{\rbrack}
\newcommand{\concat}{\mathbin{\|}}
\DeclareMathOperator*{\argmax}{arg\!\!\!\ \;max}
\newcommand{\includepdfplot}[2][]{\begingroup\includegraphics[#1]{#2}\endgroup}

\title{Tractable Pathfinding for the\texorpdfstring\\.Stochastic On-Time Arrival Problem}
\titlerunning{Tractable Pathfinding for the SOTA Problem}
\date{}
\author
{
	\href{mailto:mniknami@berkeley.edu}{Mehrdad Niknami}\inst{1} \and
	\href{mailto:samitha@cornell.edu}{Samitha Samaranayake}\inst{2}
}
\institute
{
	Electrical Engineering and Computer Science, UC Berkeley \and
	School of Civil and Environmental Engineering, Cornell University
}

\begin{document}
	\maketitle

	\begin{abstract}
		We present a new and more efficient technique for computing the route that maximizes the probability of
		on-time arrival in stochastic networks, also known as the path-based stochastic on-time arrival
		(SOTA) problem.
		Our primary contribution is a pathfinding algorithm that uses the solution to the \textit{policy}-based
		SOTA problem---which is of pseudo-polynomial-time complexity in the time budget of the journey---as
		a search heuristic for the optimal path.
		In particular, we show that this heuristic can be exceptionally efficient in practice, effectively
		making it possible to solve the path-based SOTA problem as quickly as the policy-based SOTA problem.
		Our secondary contribution is the extension of policy-based preprocessing to path-based preprocessing
		for the SOTA problem. In the process, we also introduce Arc-Potentials, a more efficient generalization
		of Stochastic Arc-Flags that can be used for \textit{both} policy- and path-based SOTA.
		After developing the pathfinding and preprocessing algorithms, we evaluate their performance
		on two different real-world networks.
		To the best of our knowledge, these techniques provide the most efficient computation strategy
		for the path-based SOTA problem for general probability distributions,
		both with and without preprocessing.
	\end{abstract}

	\section{Introduction}
		Modern advances in graph theory and empirical computational power
		have essentially rendered deterministic point-to-point routing a solved problem.
		While the ubiquity of routing and navigation tools in our everyday lives
		is a testament to the success and usefulness of deterministic routing technology,
		inaccurate predictions remain a fact of life, resulting in missed flights,
		late arrivals to meetings, and failure to meet delivery deadlines.
		Recent research in transportation engineering, therefore, has focused on
		the collection of traffic data and the incorporation of \textit{uncertainty} into traffic models,
		allowing for the optimization of relevant \textit{reliability} metrics desirable for the user.

		The point-to-point \textit{stochastic on-time arrival} problem \cite{fan2005arriving},
		or SOTA for short, concerns itself with this reliability aspect of routing.
		In the SOTA problem, the network is assumed to have uncertainty in the travel time across each link,
		represented by a strictly positive random variable.
		The objective is then to maximize the probability of on-time arrival
		when traveling between a given origin-destination pair with a fixed time budget.\footnote
		{
			The target objective can in fact be generalized to utility functions other than the probability
			of on-time arrival~\cite{blandin2014robust} with little effect on our algorithms,
			but for our purposes, we limit our discussion to this scenario.
		}
		It has been shown that the SOTA solution can appreciably increase the probability of on-time arrival compared to
		the classical \textit{least expected travel time} (LET) path~\cite{samaranayake2012tractable}, motivating the
		search for efficient solutions to this problem.

		\subsection{Variants}
		There exist two primary variants of the SOTA problem.
		The path-based SOTA problem, which is also referred to as the
		\textit{shortest-path problem with on-time arrival reliability} (SPOTAR)~\cite{nie2009shortest},
		consists of finding the a-priori most reliable path to the destination.
		The policy-based SOTA problem, on the other hand, consists of computing a \textit{routing policy}---rather
		than a fixed path---such that, at every intersection, the choice of the next direction
		depends on the current state (i.e., the remaining time budget).\footnote
		{
			In this article, we only consider time-invariant travel-time distributions.
			The problem can be extended to incorporate time-varying distributions as discussed in~\cite{samaranayake2012tractable}.
		}
		While a policy-based approach provides better reliability when online navigation is an option,
		in some situations it can be necessary to determine the entire path prior to departure.

		The policy-based SOTA problem, which is generally solved in discrete-time,
		can be solved via a successive-approximation algorithm, as shown by \citet{fan2006optimal}. This approach
		was subsequently improved by \citet{samaranayake2012tractable} to a pseudo-polynomial-time label-setting algorithm
		based on dynamic-programming with a sequence of speedup techniques and the use of
		\textit{zero-delay convolution}~\cite{dean2010speeding, samaranayake2012speedup}.
		It was then demonstrated in \citet{sabran2014precomputation} that
		graph preprocessing techniques such as Reach~\cite{gutman2004reach}
		and Arc-Flags~\cite{hilger2009fast} can be used to further reduce query times for this problem.

		In contrast with the policy-based problem, however,
		no polynomial-time solution is known for the general path-based SOTA problem~\cite{nie2009shortest}.
		In the special case of normally-distributed travel times,
		\citet{nikolova2006stochastic} present an $O(n^{O(log~n)})$-algorithm for computing exact solutions,
		while \citet{lim2013practical} present a poly-logarithmic-time algorithm for approximation solutions.
		To allow for more general probability distributions, \citet*{nie2009shortest} develop a
		label-correcting algorithm that solves the problem by utilizing the first-order stochastic dominance
		property of paths.
		While providing a solution method for general distributions, the performance of this algorithm is
		still insufficient to be of practical use in many real-world scenarios;
		for example, while the stochastic dominance approach provides a reasonable computation time
		(on the order of half a minute per instance) for networks of a few hundred to a thousand vertices,
		it fails to perform well on metropolitan road networks, which easily exceed tens of thousands of vertices.
		In contrast, our algorithm easily handles networks of tens of thousands of edges
		in approximately the same amount of time \textit{without} any kind of preprocessing.\footnote
		{
			\citet{parmentier2014stochastic} have concurrently also developed a similar approach
			concerning stochastic shortest paths with risk measures.
		}
		With preprocessing, our techniques further reduce the running time to less than
		\text{half a second}, making the problem tractable for larger networks.\footnote
		{
			It should be noted that the largest network we consider only has approximately 71,000 edges and
			is still much smaller than networks used to benchmark deterministic shortest path queries,
			which can have millions of edges~\cite{delling2009}.
		}

		\subsection{Contributions}
		Our primary contribution in this article is a practically efficient technique for solving the
		path-based SOTA problem, based on the observation that the solution to the policy-based SOTA problem
		is in practice itself an extremely efficient heuristic for solving the path-based problem.

		Our secondary contribution is to demonstrate how graph preprocessing can be used to speed up the
		computation of the policy heuristic, and thus the optimal path, while maintaining correctness.\footnote
		{
			As explained later, there is a potential pitfall that must be avoided when
			the preprocessed policy is to be used as a heuristic for the path.
		}
		Toward this goal, we present Arc-Potentials, a more efficient generalization of the existing
		preprocessing technique known as Stochastic Arc-Flags that can be used for both
		policy- and path-based preprocessing.

		After presenting these techniques, we evaluate the performance of our algorithms on two real-world
		networks while comparing the trade-off between their scalability
		(in terms of memory and computation time) and the speedups achieved.
		Our techniques, to the best of our knowledge, provide the most efficient computation strategy for the
		path-based SOTA problem with general probability distributions, both with and without preprocessing.

	\section{Preliminaries}
		We are given a stochastic network in the form of a directed graph $G = (V, E)$
		where each edge $(i, j) \in E$ has an associated probability distribution
		$p_{ij}(\cdot)$ representing the travel time across that edge.\footnote
		{
			We assume that at most one edge exists between any pair of nodes in each direction.
		}
		The source is denoted by $s \in V$, the destination by $d \in V$,
		and the travel time budget by $T \in \mathbb{R}^+$.

		For notational simplicity, we present the SOTA problem in continuous-time throughout this article,
		with the understanding that the algorithms are applied after discretization with a
		time interval of $\Delta t$.

		\begin{definition}[SOTA Policy]
			Let $u_{ij}(t)$ be the probability of arriving at the destination $d$
			with time budget $t$ when first traversing edge $(i, j) \in E$
			and subsequently following the optimal policy.
			Let $\delta_{ij} > 0$ be the minimum travel time along edge $(i, j)$, i.e. $\min \{\tau : p_{ij}(\tau) > 0\}$.
			Then, the on-time arrival probability $u_i(t)$ and the policy (optimal subsequent node) $w_i(t)$ at node $i$,
			can be defined via the dynamic programming equations below~\cite{fan2005arriving}.
			Note that $\Delta t$ must satisfy $\Delta t \leq \delta_{ij}\ \forall {(i, j) \in E}$.
			\begin{multicols}{2}
				\noindent
				\begin{align*}
					u_{ij}(t) &= \int_{\delta_{ij}}^t u_j(t - \tau) p_{ij}(\tau)\,d\tau  \\
					u_{d}(\cdot) &= 1
				\end{align*}
				\begin{align*}
					u_{i}(t) &=    \max_{j:\,(i, j) \in E} u_{ij}(t)  \\
					w_{i}(t) &= \argmax_{j:\,(i, j) \in E} u_{ij}(t)
				\end{align*}
			\end{multicols}
		\end{definition}

		The solution to the policy-based SOTA problem can be obtained by solving this system of equations
		using dynamic programming as detailed in~\cite{samaranayake2012tractable}.
		This requires evaluating a set of convolution integrals. The computation of
		the policy $u_s(\cdot)$ for a source $s$ and time budget $T$ using direct convolution takes $O(|E| T^2)$ time,
		as computing $u_s(T)$ could in the worst case require convolutions of length $O(T)$ for each edge in the graph.
		However, by using an online convolution technique
		known as \textit{zero-delay convolution} (ZDC) \cite{gardner1994efficient, dean2010speeding},
		the time complexity can be reduced to $O(|E| T \log^2 T)$.
		Justifications for the results and time complexity, including details
		on how to apply ZDC to the SOTA problem, can be found in~\cite{samaranayake2012tractable, samaranayake2012speedup}.

		\paragraph{Assumptions.}
		Our work, as with other approaches to both the policy-based and path-based SOTA problems,
		makes a number of assumptions about the nature of the travel time distributions.
		The three major assumptions are that the travel time distributions are
		(1) \textit{time invariant},
		(2) \textit{exogenous} (not impacted by individual routing choices), and
		(3) \textit{independent}.
		The time-invariance assumption---which prevents accounting for traffic variations throughout the
		day---can be relaxed under certain conditions as described in~\cite{samaranayake2012tractable}.
		Furthermore, the exogeneity assumption is made even in the case of deterministic shortest path problems.
		This leaves the independence assumption as a major concern for this problem.

		It might, in fact, be possible to partially relax this assumption~\citep{samaranayake2012tractable}
		to allow for conditional distributions at the cost of
		increasing the computation time by a factor linear in the number of states to be conditioned on.
		(If we assume the Markov property for road networks,
		the number of conditioning states becomes the in-degree of each vertex,
		a small enough constant that may make generalizations in this direction practical.)
		Nevertheless, we will only focus on the independent setting and make no claim to have
		solved the path-based SOTA problem in full generality,
		as the problem already lacks efficient solution methods even in this simplified setting.
		Our techniques should, however, provide a foundation that allows for
		relaxing these assumptions in the future.

	\section{Path-Based SOTA}

			In the deterministic setting, efficient solution strategies (from Dijkstra's
			algorithm to state-of-the-art solutions) generally exploit the \textit{sub-path optimality} property:
			namely, the fact that any optimal route to a destination node $d$ that includes some intermediate node $i$
			necessarily includes the optimal path from $i$ to $d$.
			Unfortunately, this does not hold in the stochastic setting.
			Furthermore, blind enumeration of all possible paths in the graph is
			absolutely intractable for all but the most trivial networks,
			as the number of simple paths grows exponentially with the number of nodes in the graph.
			Naturally, this leads us to seek a heuristic to guide us toward the optimal path
			efficiently, while not compromising its optimality.

		\subsection{Algorithm}

			Consider a fixed path $P$ from the source $s$ to node $i$.
			Let $q^{P}_{si}(t)$ be the travel time distribution along $P$ from node $s$ to node $i$,
			i.e., the convolution of the travel time distributions of every edge in $P$.
			Upon arriving at node $i$ at time $t$, let the user follow the optimal policy
			toward $d$, therefore reaching $d$ from $s$ with probability density
			$q^{P}_{si}(t) u_i(T - t)$.
			The reliability of following path $P$ to node $i$ and subsequently
			following the optimal policy toward $d$ is\footnote
			{
				The bounds of this integral can be slightly tightened through
				inclusion of the minimum travel times, but this has been omitted for simplicity.
			}:
			\begin{align*}
				r^{P}_{si}(T) &= \int_{0}^{T} q^{P}_{si}(t) u_i(T - t) \,dt
			\end{align*}
			Note that the route from $s \rightarrow i$ is a fixed path while that from
			$i \rightarrow d$ is a policy.

			The optimal path is found via the procedure in Algorithm~\ref{alg:SOTAPath}.
			Briefly, starting at the source $s$, we add the hybrid (path + policy)
			solution $r^{P}_{si}(T)$ for each neighbor $i$ of $s$ to a priority queue.
			Each of these solutions gives an upper bound on the solution (success probability).
			We then dequeue the solution with the highest upper bound,
			repeating this process until a path to the destination is found.

			\begin{algorithm}[t]
				\caption{Algorithm for computing the optimal SOTA path}
				\label{alg:SOTAPath}
				\begin{algorithmic}
					\State Notation:
					 $\ast$ is the convolution operator and $\concat$ is the concatenation operator
					\ForAll{$i \in V$, $0 \leq t \leq T$}
						\State Compute the optimal policy's reliability\footnotemark~$u_i(t)$
					\EndFor
					\State $Q \gets \mathrm{PriorityQueue}(\null)$
					\State $Q.\mathrm{Push}\p*{u_s(T), \p*{\bk*{1.0}, \bk*{s}}}$
						\Comment Push (reliability, (cost dist., initial path))
					\While{$Q$ is not empty}
						\State $(r, \p*{q, P}) \gets Q.\mathrm{PopMax}(\null)$
							\Comment Extract most reliable path so far
						\State $i \gets P\bk*{\abs{P} - 1}$
							\Comment Get the last node in the path
						\If{$i = d$}
							\State \Return $P$
						\EndIf
						\ForAll{$j \in E.\mathrm{Neighbors}(i)$}
							\State $Q.\mathrm{Push}\p*{\p*{q \ast u_j}\![T], \p*{q \ast E.\mathrm{Cost}(i, j), P \concat \bk*{j}}}$
								\Comment Append new edge.
						\EndFor
					\EndWhile
					\State \Return nil  \Comment No path found
				\end{algorithmic}
			\end{algorithm}
			\footnotetext
			{
				Can be limited to those $i$ and $t$ reachable from $s$ in time $T$,
				and can be further sped up through existing policy preprocessing techniques
				such as Arc-Flags.
			}
			Essentially, algorithm \ref{alg:SOTAPath} performs an $A^*$~search for the destination, using the policy
			as a heuristic.
			While it is obvious that the algorithm would find the optimal path \textit{eventually} if the search were
			allowed to continue indefinitely, it is less obvious that the \textit{first} path found will be optimal.
			We show this by showing that the policy is an admissible heuristic for the path,
			and consequently, by the optimality of $A^*$~\cite{dechter1985}, the first returned path must be optimal.

			\begin{proposition}[Admissibility]
				The solution to policy-based SOTA problem is an admissible heuristic for the
				optimal solution to the path-based SOTA problem using Algorithm \ref{alg:SOTAPath}.
			\end{proposition}
			\begin{proof}
				When finding a minimum cost path, an admissible heuristic is a heuristic that
				never overestimates the actual cost~\cite{russell1995}.
				In our context, since the goal is to maximize the reliability (success probability),
				this corresponds to a heuristic that never underestimates the reliability of a
				routing strategy. The reliability of an optimal SOTA policy clearly provides an upper bound
				on the reliability of any fixed path with the same source, destination, and travel budget.
				(Otherwise, a better policy would be to simply follow the fixed path irrespective of
				the time remaining, contradicting the assumption that the policy is optimal.)
				Therefore, the SOTA policy is an admissible heuristic for the optimal SOTA path.
			\end{proof}

		\subsection{Analysis}

			The single dominant factor in this algorithm's (in)efficiency is the
			length of the priority queue (i.e., the number of paths considered by the algorithm),
			which in turn depends on the travel time distribution along each road.
			As long as the number of paths considered is approximately linear in length of the optimal path,
			the path computation time is easily dominated by the policy computation time, and the algorithm
			finds the optimal path very quickly.
			In the worst-case scenario for the algorithm, the optimal path at a node corresponds to the
			direction for the worst policy at that node.
			Such a scenario, or even one in which the optimal policy frequently chooses a suboptimal path,
			could result in a large (even exponential) running time as well as space usage.
			However, it is difficult to imagine this happening in practice.
			As shown later, experimentally, we came across very few cases in which the path
			computation time dominated the policy computation time, and even in those cases, they were
			still quite reasonable and extremely far from such a worst-case scenario.
			We conjecture that such situations are extremely unlikely to occur in real-world road networks.

			An interesting open problem is to define characteristics (network structure, shape of distributions, etc.)
			that guarantee pseudo-polynomial running times in stochastic networks,
			similar in nature to the \textit{Highway Dimension} property~\cite{abraham2010highway}
			in deterministic networks,
			which guarantees logarithmic query times when networks have a low \textit{Highway Dimension}.

	\section{Preprocessing}

		In deterministic pathfinding, preprocessing techniques such as
		\textit{Arc-Flags} \citep{hilger2009fast},
		\textit{reach-based routing} \citep{gutman2004reach,goldberg2006reach},
		\textit{contraction hierarchies} \citep{geisberger2008contraction}, and
		\textit{transit node routing} \citep{bast2006transit}
		have been very successfully used to decrease query times by many orders of magnitude by exploiting
		structural properties of road networks.
		Some of these approaches allow for pruning the search space based solely on the destination node,
		while others also take the source node into account, allowing for better pruning at the cost of
		additional preprocessing. The structure of the SOTA problem, however, makes it more challenging to apply
		such techniques to it.
		Previously, Arc-Flags and Reach have been successfully adapted to the policy-based problem
		in~\citep*{sabran2014precomputation}, resulting in Stochastic Arc-Flags and Directed Reach.
		While at first glance one may be tempted to directly apply these algorithms to the computation of the
		policy heuristic for the path-based problem,
		a naive application of source-dependent pruning (such as Directed Reach or source-based Arc-Flags)
		can result in an incorrect solution,
		as the policy needs to be recomputed for source nodes that correspond to different source regions.
		This effectively limits any preprocessing of the policy heuristic to destination-based
		(i.e., source-independent) techniques such as Stochastic Arc-Flags, precluding the use of source-based
		approaches such as Directed Reach for the policy computation.

		With sufficient preprocessing resources (as explained in section~\ref{sec:scalability}),
		however, one can improve on this through the direct use of \textit{path-based} preprocessing---that is,
		pruning the graph to include only those edges which may be part of the most reliable path.
		This method allows us to simultaneously account for both source and destination regions, and generally
		results in a substantial reduction of the search space on which the policy needs to be computed.
		However, as path-based approaches require computing paths between all $\approx\abs{V}^2$ pairs of vertices
		in the graph, this approach may become computationally prohibitive for medium- to large-scale networks.
		In such cases, we would then need to either find alternate approaches (e.g. approximation techniques),
		or otherwise fall back to the less aggressive policy-based pruning techniques,
		which only require computing $\abs{V}$ separate policies (one per destination).

	\subsection{Efficient Path-based Preprocessing}
		Path-based preprocessing requires finding the optimal paths for each time budget up to the desired time
		budget $T$ for all source-destination pairs.
		Naively, this can be done by placing Algorithm~\ref{alg:SOTAPath} in a loop,
		executing it for all time budgets from $1$ to $T$.
		This requires $T$ times the work of finding the path for a single time budget,
		which is clearly prohibitive for any reasonable value of $T$.
		However, we can do far better by observing that many of the computations in the algorithm are
		independent of the time budget and can be factored out when the path does not change with $T$.

		To improve the efficiency of the naive approach in this manner, we make two observations.
		First, we observe that, in Algorithm~\ref{alg:SOTAPath}, \textit{only} the computation of
		the path's reliability (priority) in the priority queue ever requires knowledge of the time budget.
		Crucially, the convolution $q \ast E.\mathrm{Cost}(i, j)$ only depends on the \textit{maximum}
		time budget $T$ for truncation purposes, which is a fixed value.
		This means that the travel time distribution of any path under consideration can be computed
		once for the maximum time budget, and re-used for all lower time budgets thereafter.
		Second, we observe that when a new edge is appended, the priority of the new path is the inner product
		of the vector $q$ and (the reverse of) the vector $u_j$, shifted by $T$.
		As noted in the algorithm itself, this quantity in fact the convolution
		of the two aforementioned vectors evaluated at $T$.
		Thus, when a new edge is appended, instead of recomputing the inner product, we can simply
		convolve the two vectors once, and thereafter look up the results instantly for other time budgets.

		Together, these two observations allow us to compute the optimal paths for all budgets far faster than
		would seem naively possible, making path-based preprocessing a practical option.

	\subsection{Arc-Potentials}
		As noted earlier, Arc-Flags, a popular method for graph preprocessing,
		has been adapted to the SOTA problem as Stochastic Arc-Flags \cite*{sabran2014precomputation}.
		Instead of applying it directly, however, we present \textit{Arc-Potentials}, a more
		natural generalization of Arc-Flags to SOTA that can still be directly applied to the policy- and
		path-based SOTA problems alike, while allowing for more efficient preprocessing.

		Consider partitioning the graph $G$ into $R$ regions (we choose $R = O(\log |E|)$, described below),
		where $R$ is tuned to trade off storage space for pruning accuracy.
		In the deterministic setting, Arc-Flags allow us to preprocess and prune the search space as follows.
		For every arc (edge) $(i, j) \in E$, Arc-Flags defines a bit-vector of length $R$ that denotes whether
		or not this arc belongs to an optimal path ending at some node in region $R$.
		We then pre-compute these Arc-Flags, and store them for pruning the graph at query time.
		(This approach has been extended to the dynamic setting \cite{d2011dynamic}
		in which the flags are updated with low recomputation cost after the road network is changed.)

		\citet{sabran2014precomputation} apply Arc-Flags to the policy-based SOTA problem as follows:
		each bit vector is defined to represent whether or not its associated arc is \textit{realizable},
		meaning that it \textit{belongs to an optimal policy} to some destination
		in the target region associated with each bit.
		The problem with this approach, however, is that it requires computing arc-flags for \textit{all}
		target budgets (or more, practically, some ranges of budgets), each of which takes a considerable amount of space.
		Instead, we propose a more efficient alternative \ref{def:ArcPotentials}, which we call \textit{Arc-Potentials}.
		\begin{definition}[Arc-Potentials]
			\label{def:ArcPotentials}
			For a given destination region $D$,
			we define the \textit{arc activation potential} $\phi_{ij}$ of the edge from
			node $i$ to node $j$ to be the
			\textit{minimum} time budget at which the arc becomes part of an optimal policy to some
			destination $d \in D$.
		\end{definition}

		The Arc-Potentials pruning algorithm only stores the \textit{``activation''} potential of every edge.
		As expected, this implies that for large time budgets, every edge is potentially active.
		We could have further generalized the algorithm to allow for asymptotically \textit{exact}
		pruning at relatively low cost by simply storing the actual potential \textit{intervals}
		during which the arc is active, rather than merely their first activation potential.
		However, in our experiments this was deemed unnecessary as Arc-Potentials were already
		sufficient for significant pruning in the time budgets of interest in our networks.

		The computation of the set of realizable edges (and nodes) under a given policy
		is essentially equivalent to the computation of the policy itself,
		except that updates are performed in the reverse order (from the source to the destination).
		The activation potentials $\phi$ can then be obtained from this set.
		As with Arc-Flags, we limit the space complexity to $O(|E| R) = O(|E| \log |E|)$
		by choosing $R$ to be proportional $\log|E|$,
		tuning it as desired to increase the pruning accuracy.
		In our experiments, we simply used a rectangular grid of size $\sqrt{R}\times \sqrt{R}$.
		Note, however, that the preprocessing time does not depend on $R$,
		as the paths between all $\approx\abs{V}^2$ pairs of nodes must be eventually computed.

	\section{Experimental Results}

			We evaluated the performance of our algorithms on two real-world test networks: a small San~Francisco
			network with $2643$ nodes and $6588$ edges for which real-world travel-time data was
			available as a Gaussian mixture model~\citep{hunter2013path}, and a second (relatively larger)
			Luxembourg network with $30647$ nodes and $71655$ edges for which travel-time distributions
			were synthesized from road speed limits, as real-world data was unavailable.
			The algorithms were implemented in C++ (2003) and executed on a cluster of
			$1.9$~GHz AMD~Opteron\texttrademark~6168 CPUs. The SOTA queries were executed on a single CPU
			and the preprocessing was performed in parallel as explained below.

			The SOTA policies were computed as explained in~\cite{samaranayake2012tractable,samaranayake2012speedup}
			using zero-delay convolution with a discretization interval of
			$\Delta t = 1\ \mathrm{second}$.\footnote
			{
				Recall that we must have $\Delta t \leq \min_{(i,j)\in E} \delta_{ij}$, which is
				$\approx 1~\mathrm{sec}$ for our networks.
			}
			To generate random problem instances, we independently picked a source and a destination node
			uniformly at random from the graph
			and computed the least expected travel-time (LET) path between them.
			We then evaluated our pathfinding algorithm
			for budgets chosen uniformly at random from the $5^\mathrm{th}$ to $95^\mathrm{th}$
			percentile of LET~path travel times (those of practical interest) on
			$10,000$ San~Francisco and $1000$ Luxembourg problems instances.

			First, we discuss the speed of our pathfinding algorithm, and afterward,
			we evaluate the effectiveness and scalability of our preprocessing strategies.

		\subsection{Evaluation}

\begin{figure}[h]
				\centering
				\begin{subfigure}{0.48\textwidth}
					\centering
					\includegraphics[width=\textwidth]{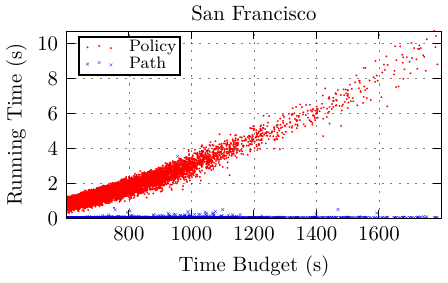}
				\end{subfigure}
				\begin{subfigure}{0.448\textwidth}
					\centering
					\includegraphics[width=\textwidth]{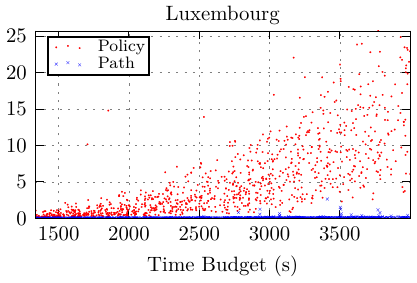}
				\end{subfigure}
				\caption
				{
					Running time of the pathfinding algorithm as a function of the travel time
					budget for random unpruned
					(i.e., non-preprocessed) instantiations of each network.
					We can see that the path computation time is dominated
					by the policy computation time, effectively
					reducing the path-based SOTA problem to the policy-based SOTA problem
					in terms of computation time.
				}
				\label{fig:Pathfinding-ZDC}
			\end{figure}

			We first evaluate the performance of our path-based SOTA algorithm without any graph preprocessing.
			Experimental results, as can be seen in Figure~\ref{fig:Pathfinding-ZDC}, show that the run time of
			our solution is dominated by the time taken to obtain the solution to the policy-based SOTA problem,
			which functions as a search heuristic for the optimal path.

			The stochastic-dominance (SD) approach \cite*{nie2009shortest},
			which to our knowledge is the fastest published solution
			for the path-based SOTA problem with general probability distributions,
			takes, on average, between $7$ and $18$ seconds
			(depending on the variance of the link travel time distributions)
			to compute the optimal path for $100$~time-step budgets.
			For comparison, our algorithm solves for paths on the San~Francisco network with
			budgets of up to $1400$~seconds (= $1400$ time-steps) in $\approx 7$ seconds, even
			achieving query times below $1$~second for budgets less than $550$~seconds
			without any preprocessing at all.
			Furthermore, it also handles most queries on the $71655$-edge Luxembourg
			network in $\approx 10$~seconds (almost all of them in $20$~seconds),
			where the network and time budgets are more than an order of magnitude larger than the
			$2950$-edge network handled by the SD approach in the same amount of time.

			Of course, this speedup---which increases more dramatically with the problem size---is
			hardly surprising or coincidental; indeed, it is quite fundamental to the nature of the algorithm:
			by drawing on the optimal policy as an upper bound (and quite often an accurate one) for the reliability
			of the final path, it has a very clear and fundamental informational advantage over any search algorithm
			that lacks any knowledge of the final solution.
			This allows the algorithm to direct itself toward the final path in an intelligent manner.

			It is, however, less clear and more difficult to see how one might compare the
			performance of our generic discrete-time approach with Gaussian-restricted,
			continuous-time approaches~\cite{lim2013practical,lim2010stochastic}.
			Such approaches operate under drastically different assumptions and,
			in the case of \citep{lim2013practical}, use approximation techniques,
			which we have yet to employ for additional performance improvements.
			When the true travel times cannot be assumed to follow Gaussian distributions,
			however, our method, to the best of our knowledge, presents the most efficient means
			for solving the path-based SOTA problem.

			As we show next, combining our algorithm with preprocessing techniques allows us to achieve
			even further reductions in query time, making it more tractable for industrial applications on
			appropriately sized networks.

\begin{figure}[H]
				\centering
				\begin{subfigure}{\textwidth}
					\centering
					\begin{subfigure}{0.32\textwidth}
						\centering
						\includegraphics[width=\textwidth]{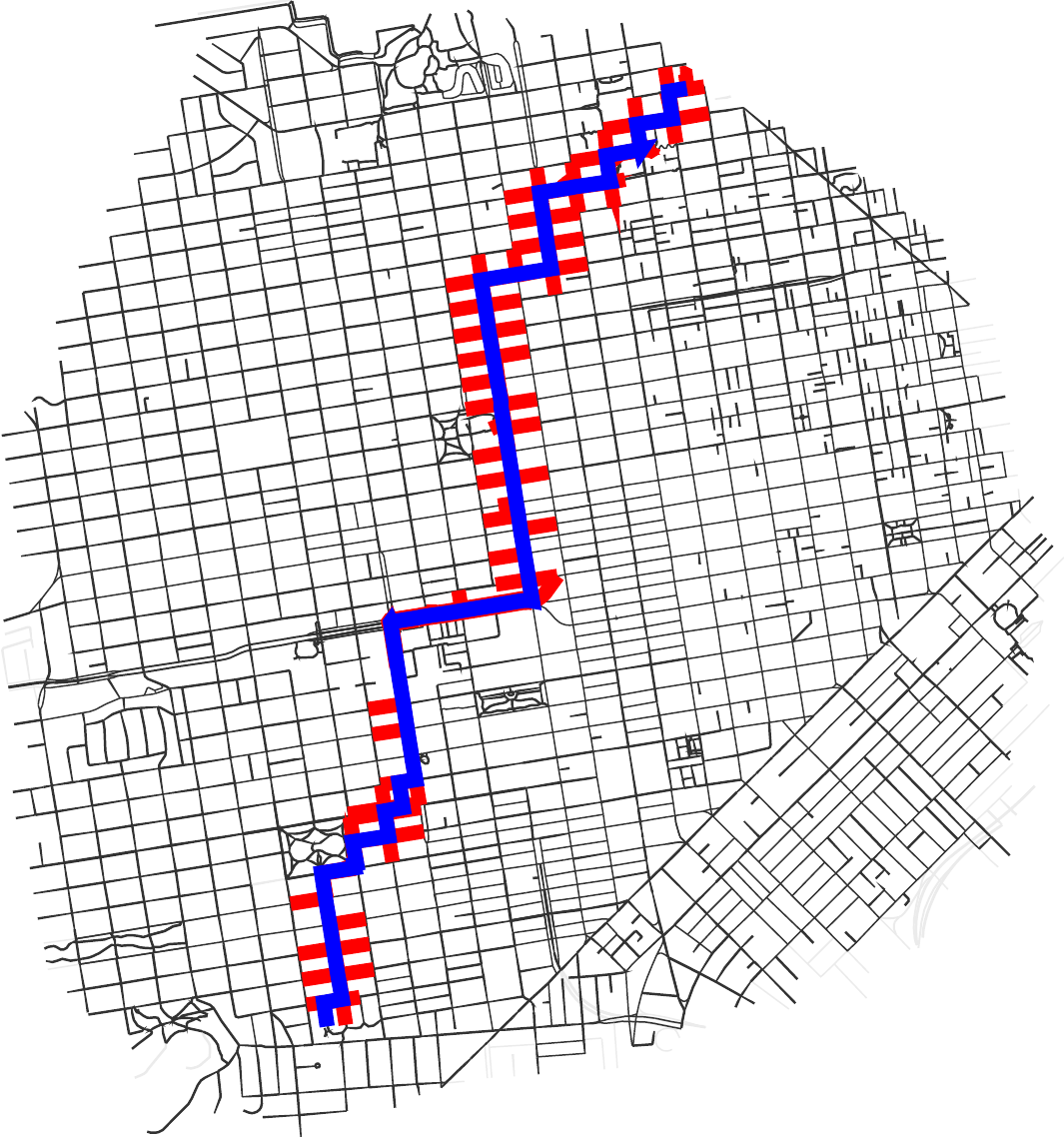}
						\caption*{$0$ regions (unpruned)}
					\end{subfigure}
					\begin{subfigure}{0.32\textwidth}
						\centering
						\includegraphics[width=\textwidth]{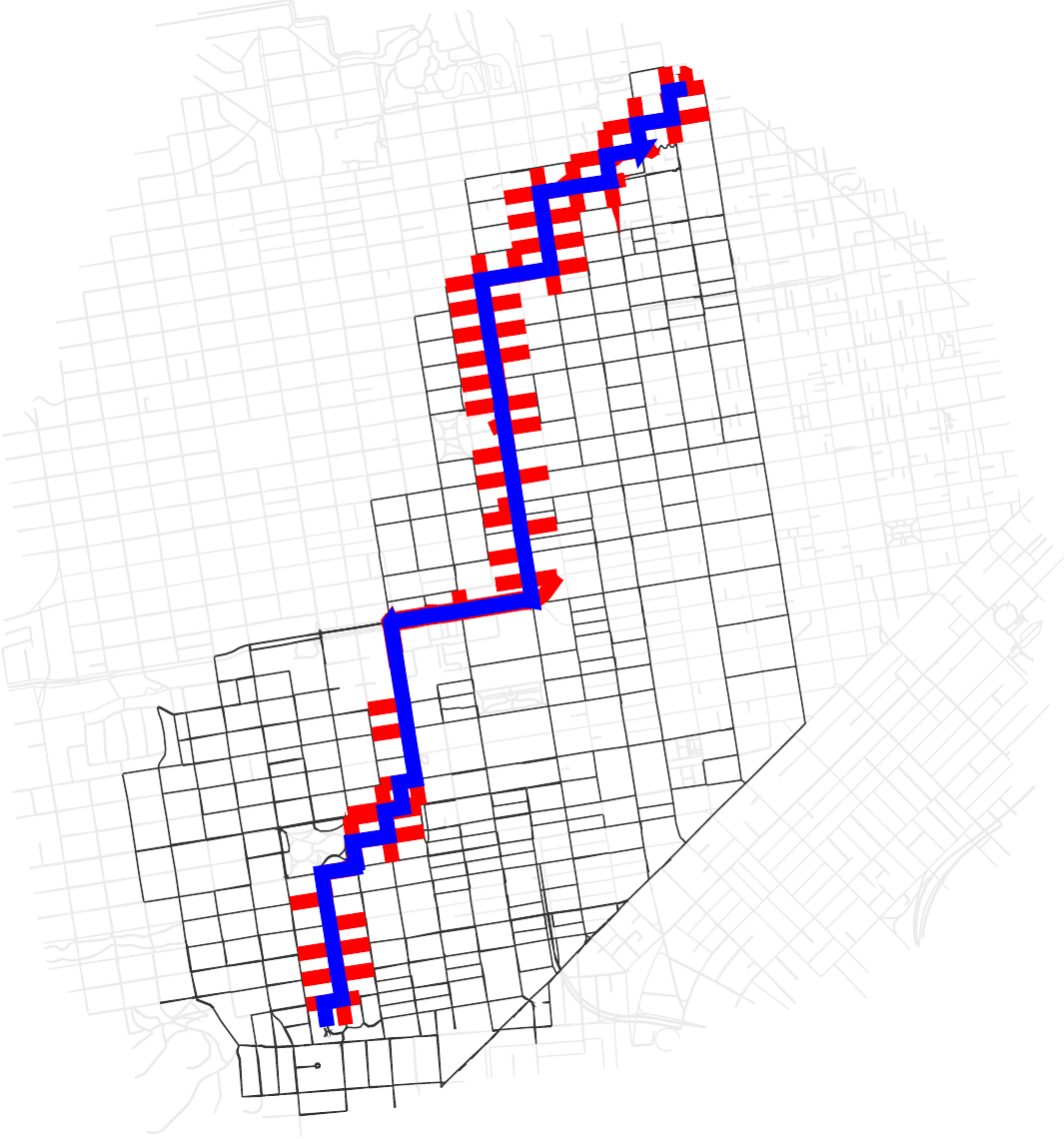}
						\caption*{$13^2$ regions (policy-pruned)}
					\end{subfigure}
					\begin{subfigure}{0.32\textwidth}
						\centering
						\includegraphics[width=\textwidth]{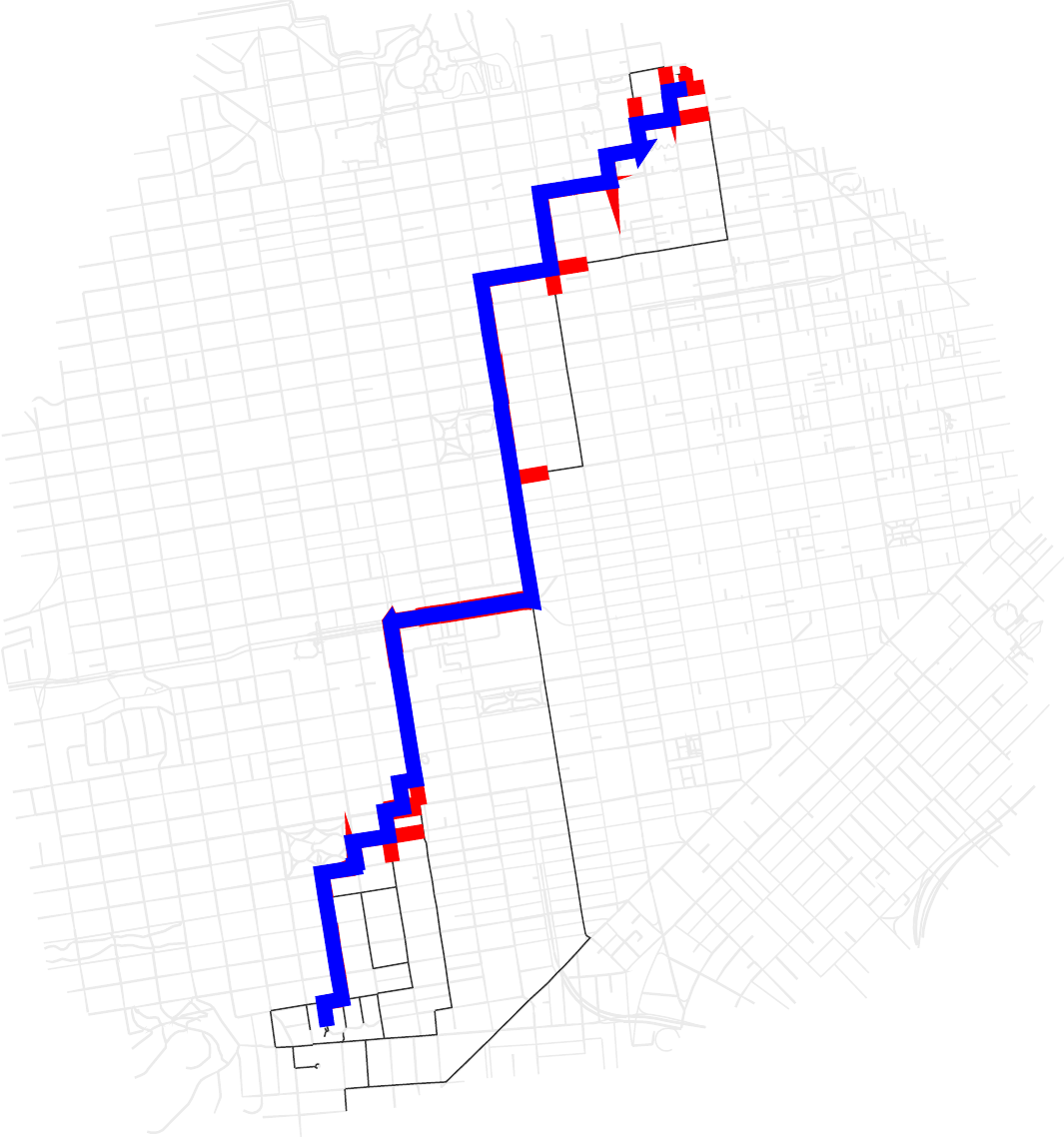}
						\caption*{$10^2$ regions (path-pruned)}
					\end{subfigure}
				\end{subfigure}
				\begin{subfigure}{\textwidth}
					\centering
					\begin{subfigure}{0.32\textwidth}
						\centering
						\includegraphics[width=\textwidth]{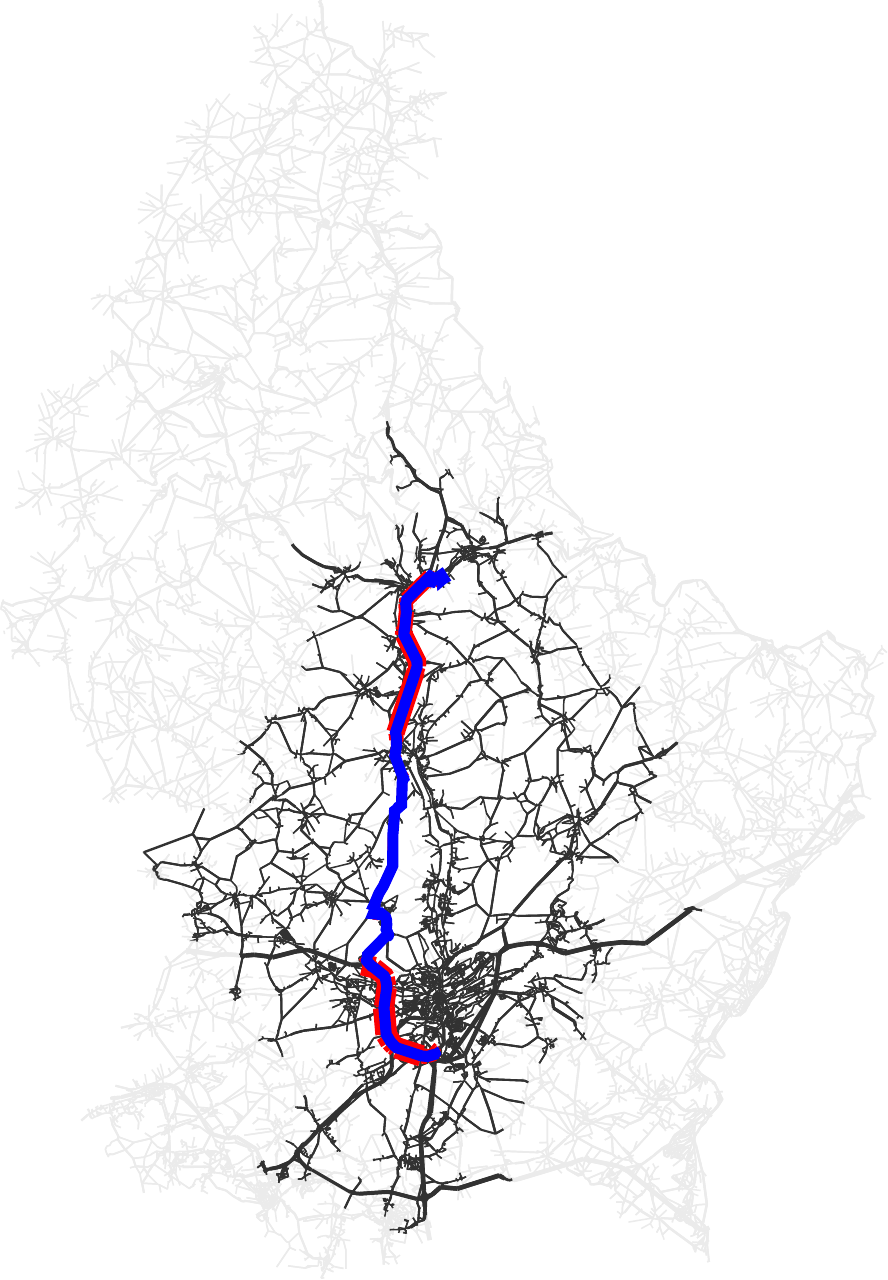}
						\caption*{$0$ regions (unpruned)}
					\end{subfigure}
					\begin{subfigure}{0.32\textwidth}
						\centering
						\includegraphics[width=\textwidth]{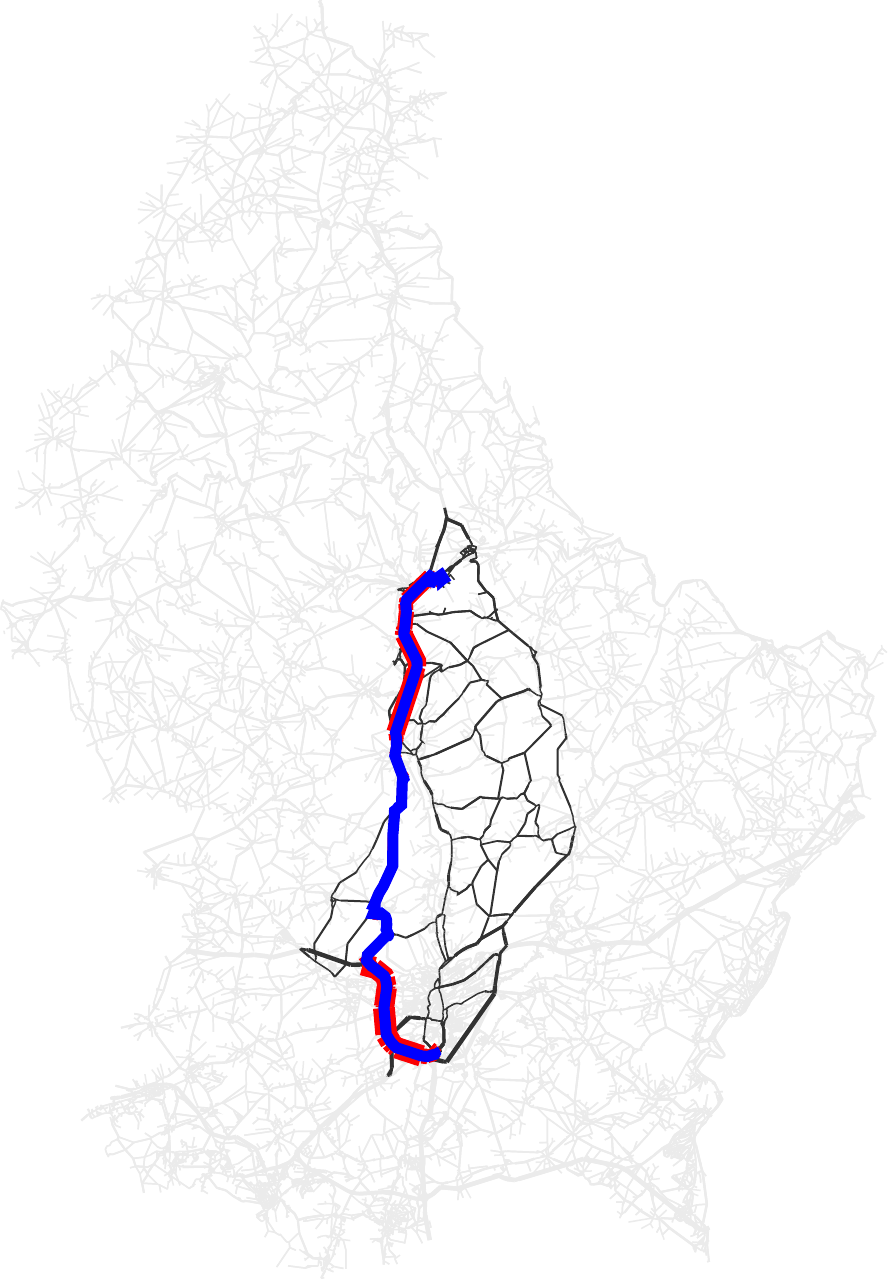}
						\caption*{$23^2$ grid (policy-pruned)}
					\end{subfigure}
					\begin{subfigure}{0.32\textwidth}
						\centering
						\includegraphics[width=\textwidth]{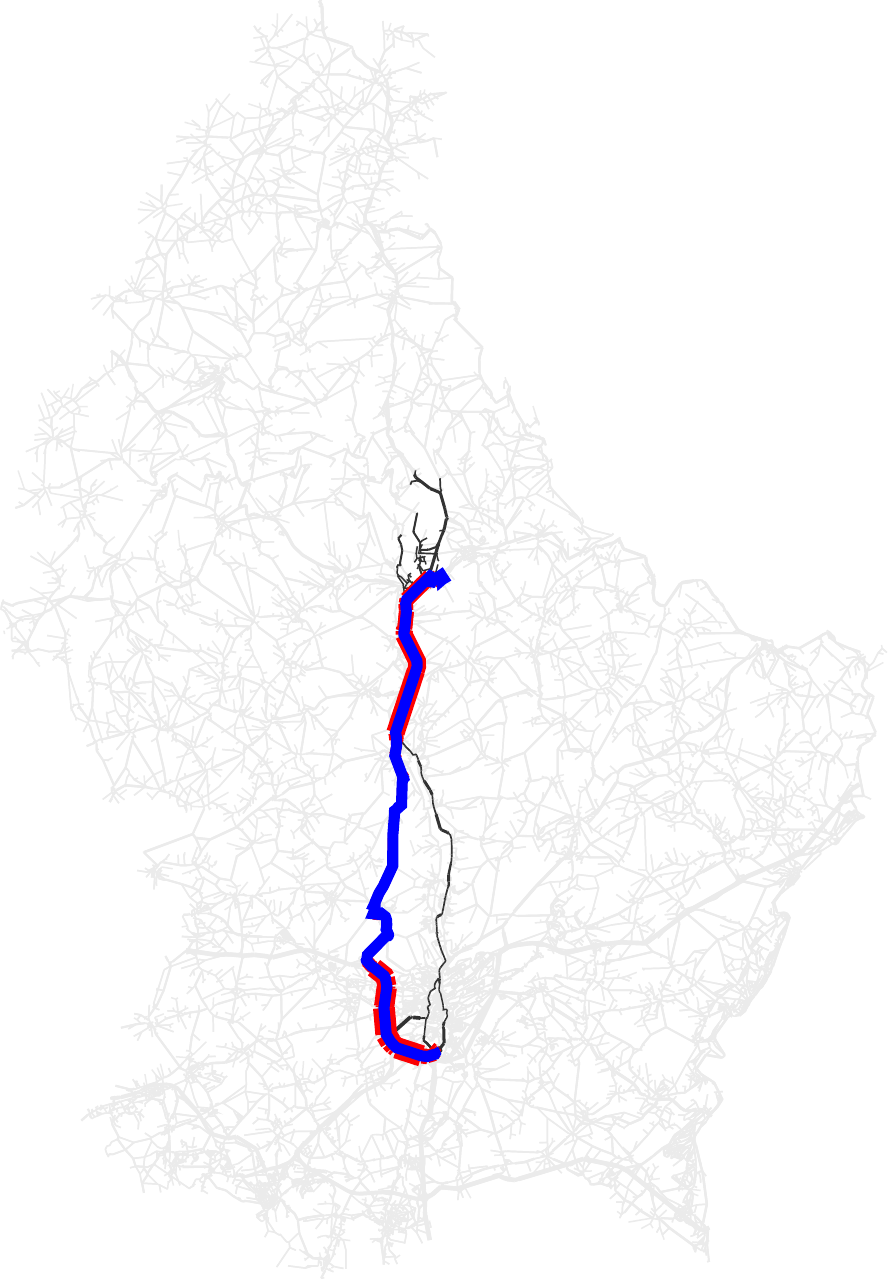}
						\caption*{$17^2$ regions (path-pruned)}
					\end{subfigure}
				\end{subfigure}
				\caption
				{
					Policy- vs. path-based pruning for random instances of
					San~Francisco ($T = 837$~seconds, source at top)
					and Luxembourg ($T = 3165$~seconds, source at bottom).
					Light-gray edges are pruned from the graph and blue edges belong to the optimal
					path, whereas red edges belong to (sub-optimal) paths that were on the queue
					at the termination of the algorithm.
				}
				\label{fig:Path}
			\end{figure}

		\paragraph{Preprocessing.}
			Figure \ref{fig:Path} demonstrates policy-based and path-based preprocessing using Arc-Potentials
			for two random San~Francisco and Luxembourg problem instances.
			As can be seen in the figure, path-based preprocessing is in general much more effective
			than policy-based preprocessing.

\begin{figure}[H]
				\centering
				\begin{subfigure}[t]{0.512\textwidth}
					\centering
					\includepdfplot[width=\textwidth]{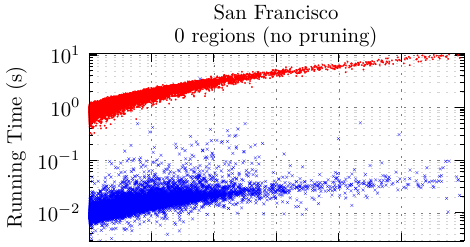}
					\includepdfplot[width=\textwidth]{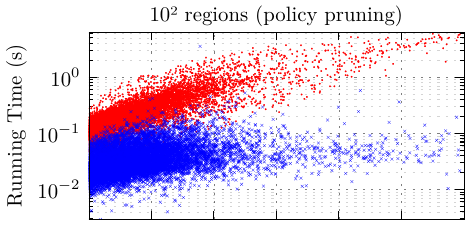}
					\includepdfplot[width=\textwidth]{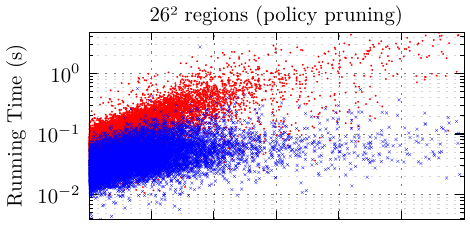}
					\includepdfplot[width=\textwidth]{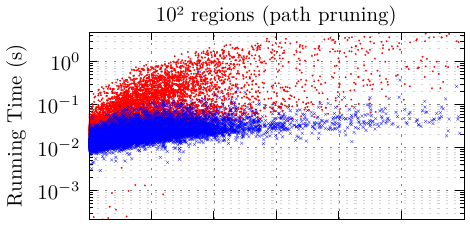}
					\includepdfplot[width=\textwidth]{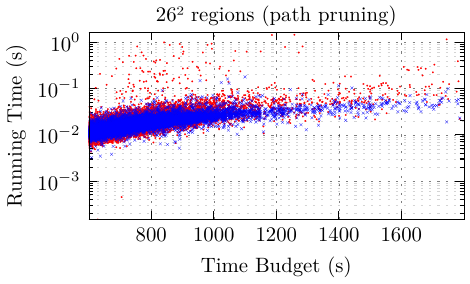}
				\end{subfigure}
				\begin{subfigure}[t]{0.478\textwidth}
					\centering
					\includepdfplot[width=\textwidth]{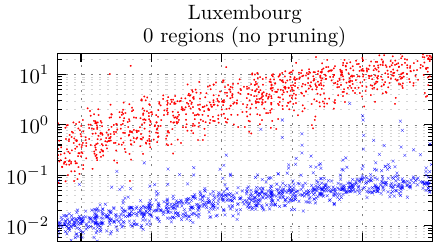}
					\includepdfplot[width=\textwidth]{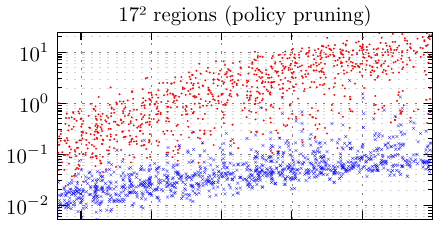}
					\includepdfplot[width=\textwidth]{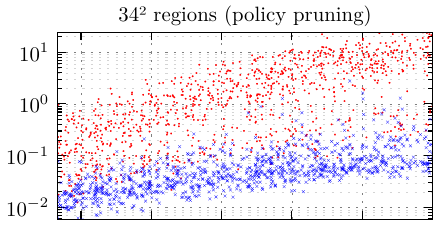}
					\includepdfplot[width=\textwidth]{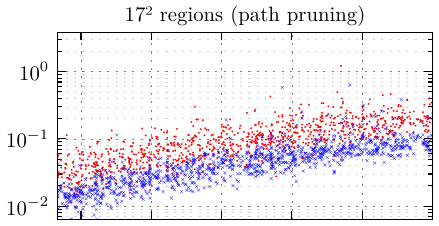}
					\includepdfplot[width=\textwidth]{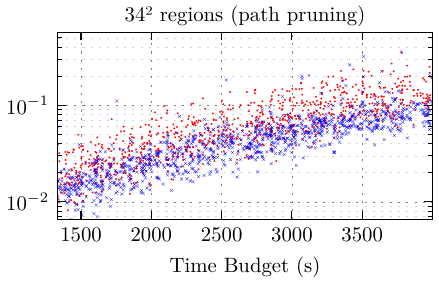}
				\end{subfigure}
				\caption
				{
					Running time of pathfinding algorithm as a function of the time budget for each
					network. Red dots represent the computation time of the policy,
					and blue crosses represent the computation of the path using that policy.
				}
				\label{ref:Policy_Against_Path-ZDC}
			\end{figure}

			Figure \ref{ref:Policy_Against_Path-ZDC},
			summarized in table \ref{ref:Policy_Against_Path-ZDC-Table},
			shows how the computation times scale with the preprocessing parameters.
			As expected, path-based preprocessing performs much better than purely policy-based preprocessing,
			and both become faster as we use more fine-grained regions.
			Nevertheless, we see that the majority of the speedup is achieved via a small number of regions,
			implying that preprocessing can be very effective even with low amounts of storage.
			(For example, for a $17\times 17$ grid in Luxembourg, this amounts to
			$71655 \times 17^2 \approx 21\mathrm{M}$ floats.)

			\begin{table}[H]
				\centering
				\caption
				{
					The average query time with both policy-based and path-based pruning
					at various grid sizes and time budgets on the San~Francisco network (left)
					and the Luxembourg network (right).
					We can see that in both cases, most of the speedup occurs at low granularity
					(and thus low space requirements).
				}
				\begin{tabular}{|c|c|c|c|c|c|} \noalign{\smallskip}\hline
					Grid/ & \multicolumn{5}{c|}{Time budget (seconds)}  \\\cline{2-6}
					Pruning & 800 & 1000 & 1200 & 1400 & 1600     \\
					\hline
					Unpruned        & 1.81 & 3.00 & 4.10 & 5.11 & 5.23  \\ \hline
					$10\times 10$, policy & 0.30 & 0.69 & 1.11 & 1.66 & 1.72  \\ \hline
					$26\times 26$, policy & 0.17 & 0.40 & 0.63 & 0.93 & 0.97  \\ \hline
					$10\times 10$, path   & 0.11 & 0.38 & 0.63 & 0.87 & 0.90  \\ \hline
					$26\times 26$, path   & 0.02 & 0.04 & 0.06 & 0.07 & 0.08  \\ \hline
				\end{tabular}
				\begin{tabular}{|c|c|c|c|c|c|} \noalign{\smallskip}\hline
					Grid/ & \multicolumn{5}{c|}{Time budget (seconds)}  \\\cline{2-6}
					Pruning & 1500 & 2000 & 2500 & 3000 & 3500     \\
					\hline
					Unpruned        & 0.54 & 1.29 & 3.53 & 6.27 & 9.95  \\ \hline
					$17\times 17$, policy & 0.25 & 0.83 & 2.31 & 4.57 & 7.97  \\ \hline
					$34\times 34$, policy & 0.21 & 0.71 & 2.09 & 3.79 & 7.07  \\ \hline
					$17\times 17$, path   & 0.03 & 0.06 & 0.09 & 0.13 & 0.18  \\ \hline
					$34\times 34$, path   & 0.02 & 0.04 & 0.06 & 0.08 & 0.12  \\ \hline
				\end{tabular}
				\label{ref:Policy_Against_Path-ZDC-Table}
			\end{table}

		\subsection{Scalability}\label{sec:scalability}

			\textit{Path}-based preprocessing requires routing between all $\approx\abs{V}^2$ pairs of vertices,
			which is quadratic in the size of the network and intractable for moderate size networks.
			In practice, this meant that we had to preprocess every region lazily (i.e. on-demand),
			which on our CPUs took $9000$ CPU-hours.
			It is therefore obvious that this becomes intractable for large networks, leaving
			policy-based preprocessing as the only option.
			One possible approach for large-scale path-based preprocessing might be to
			consider the boundary of each region rather than its interior \cite{sabran2014precomputation}.
			While currently uninvestigated, such techniques may prove to be extremely useful in practice,
			and are potentially fruitful topics for future exploration.

	\section{Conclusion and Future Work}

		We have presented an algorithm for solving the path-based SOTA problem by first
		solving the easier policy-based SOTA problem and then using its solution as a search heuristic.
		We have also presented two approaches for preprocessing the underlying network to speed up computation of
		the search heuristic and path query,
		including a generalization of the Arc-Flags preprocessing algorithm that we call Arc-Potentials.
		We have furthermore applied and implemented these algorithms on moderate-sized transportation
		networks and demonstrated their potential for high efficiency in real-world networks.

		While unobserved in practice, there remains the possibility that our algorithm
		may perform poorly on stochastic networks
		in which the optimal policy is a poor heuristic for the path reliability.
		Proofs in this direction have remained elusive, and determining whether
		such scenarios can occur in realistic networks remains an important step for future research.
		In the absence of theoretical advances, however, our algorithm provides a more tractable
		alternative to the state-of-the-art techniques for solving the path-based SOTA problem.

		While our approach is tractable for larger
		networks than were possible with previous solutions, it does not scale well enough
		to be used with regional or continental sized networks that modern deterministic shortest
		path algorithms can handle with ease.
		In the future, we hope to investigate how our policy-based approach
		might be combined with other techniques such as
		the first-order stochastic dominance~\cite{nie2009shortest} and approximation methods such
		approximate Arc-Flags~\cite{sabran2014precomputation} for further speedup,
		and to also look into algorithms that allow for at least a partial relaxation of the
		independence assumption.
		We hope that our techniques will provide a strong basis for even better algorithms
		to tackle this problem for large-scale networks in the foreseeable future.

	\bibliographystyle{unsrtnat}
	\bibliography{\jobname}
\end{document}